\documentclass[%
 aip,
 amsmath,amssymb,
 reprint,%
]{revtex4-2}
\usepackage{ulem}
\usepackage{graphicx}
\usepackage{dcolumn}
\usepackage{bm}
\usepackage{color}
\usepackage{xcolor}
\usepackage[utf8]{inputenc}
\usepackage[T1]{fontenc}
\usepackage{mathptmx}
\usepackage{etoolbox}
\usepackage{tabularx}
\usepackage{lineno}
\usepackage{soul}

\makeatletter
\def\@email#1#2{%
 \endgroup
 \patchcmd{\titleblock@produce}
  {\frontmatter@RRAPformat}
  {\frontmatter@RRAPformat{\produce@RRAP{*#1\href{mailto:#2}{#2}}}\frontmatter@RRAPformat}
  {}{}
}%
\makeatother
\begin{document}

\preprint{AIP/123-QED}

\title[]{Shubnikov-de Haas oscillations of biaxial-strain-tuned superconductors in pulsed magnetic field up to 60~T}
\author{King Yau Yip$^\dag$}
\author{Lingfei Wang$^\dag$}
\author{Tsz Fung Poon}
\author{Kai Ham Yu}
\author{Siu Tung Lam}
\affiliation{Department of Physics, The Chinese University of Hong Kong, Shatin, Hong Kong, China}
\author{Kwing To Lai}
\affiliation{Department of Physics, The Chinese University of Hong Kong, Shatin, Hong Kong, China}
\affiliation{Shenzhen Research Institute, The Chinese University of Hong Kong, Shatin, Hong Kong, China}
\author{John Singleton}
\author{Fedor F. Balakirev$^*$}
\affiliation{National High Magnetic Field Laboratory, Los Alamos National Laboratory, Los Alamos, New Mexico 87545, USA}
\author{Swee K. Goh$^*$}
\email{fedor@lanl.gov and skgoh@cuhk.edu.hk}
\affiliation{Department of Physics, The Chinese University of Hong Kong, Shatin, Hong Kong, China}

\date{\today}

\begin{abstract}
Two-dimensional (2D) materials have gained increasing prominence not only in fundamental research but also in daily applications. However, to fully harness their potential, it is crucial to optimize their properties with an external parameter and track the electronic structure simultaneously. Magnetotransport over a wide magnetic field range is a powerful method to probe the electronic structure and, for metallic 2D materials, quantum oscillations superimposed on the transport signals encode Fermi surface parameters. In this manuscript, we utilize biaxial strain as an external tuning parameter and investigate the effects of strain on the electronic properties of two quasi-2D superconductors, MoTe$_2$ and RbV$_3$Sb$_5$, by measuring their magnetoresistance in pulsed magnetic fields up to 60~T. With a careful selection of insulating substrates, we demonstrate the possibility of both the compressive and tensile biaxial strain, imposed on MoTe$_2$ and RbV$_3$Sb$_5$, respectively. For both systems, the applied strain has led to superconducting critical temperature enhancement compared to their free-standing counterparts, proving the effectiveness of this biaxial strain method at cryogenic temperatures. Clear quantum oscillations in the magnetoresistance -- the Shubnikov-de Haas (SdH) effect -- are obtained in both samples. In strained MoTe$_2$, the magnetoresistance exhibits a nearly quadratic dependence on the magnetic field and remains non-saturating even at the highest field. Whereas in strained RbV$_3$Sb$_5$, two SdH frequencies showed a substantial enhancement in effective mass values, hinting at a possible enhancement of charge fluctuations. Our results demonstrate that combining biaxial strain and pulsed magnetic field paves the way for studying 2D materials under unprecedented conditions.


\end{abstract}

\maketitle

Rapid progress in the discovery, synthesis, and reliable exfoliation of two-dimensional (2D) or quasi-2D materials has revolutionized our daily lives by offering an extensive range of applications. Notably, graphene-related devices have played a critical role in rechargeable battery technology where energy efficiency has been greatly enhanced~\cite{pumera2009,bonaccorso2015,lv2016,lavagna2020}; the high transparency and mechanical flexibility exhibited by transition metal dichalcogenides (TMD) have positioned them as promising candidates for next-generation solar cells~\cite{bernardi2013,akama2017,sumesh2019,sulas2020}. Furthermore, emerging fields based on 2D materials, such as ``twistronics''~\cite{Carr2017,Ciarrocchi2022}, also open up new avenues for harnessing exotic physical phenomena. Given the prominence of 2D materials in modern condensed matter research, it is necessary to thoroughly understand and control their electronic structure.

A powerful means to modify the electronic properties of bulk, three-dimensional (3D) materials is to subject them to hydrostatic pressure~\cite{Jayaraman1983,Miletich2000,Shen2016}. One can then contemplate an equivalent approach for tuning 2D materials: since hydrostatic pressure shrinks the volume of 3D materials uniformly, an effective methodology is needed to vary the area of the 2D materials uniformly. Thus, an in-plane, biaxial strain is an effective tuning parameter for 2D materials, as has been demonstrated in Fe-based superconductors~\cite{Bohmer2017,Nakajima2021}.

The modification of the sample volume or area changes the interatomic spacing. Consequently, the electronic structure can be altered. Indeed, not many materials come immediately with desirable properties without suitable optimisations. With the modification of the electronic structure, it is then necessary to determine the carrier densities and mobilities, and for metallic 2D materials, the Fermi surfaces.

The magnetic field is extremely useful for probing electronic properties. First, measuring electrical transport in a magnetic field enables the decoupling of the carrier mobility and density, which would otherwise be impossible to determine independently. Next, the Shubnikov-de Haas (SdH) effect, magnetic quantum oscillations in electrical resistance, is a powerful tool for Fermi surface characterization~\cite{Shoenberg2009}. Generally, the higher the field strength, the more pronounced the quantum oscillations because the Landau levels are less broadened. Thus, the ability to tune 2D materials with biaxial strain, and then conduct magnetotransport measurements at high fields, will be beneficial for investigating a variety of 2D materials. In this manuscript, we demonstrate the ability to detect the SdH effect in superconducting MoTe$_2$ and RbV$_3$Sb$_5$ under biaxial strain in a pulsed magnet.

\begin{figure}[!t]\centering
      \resizebox{8.5cm}{!}{
              \includegraphics{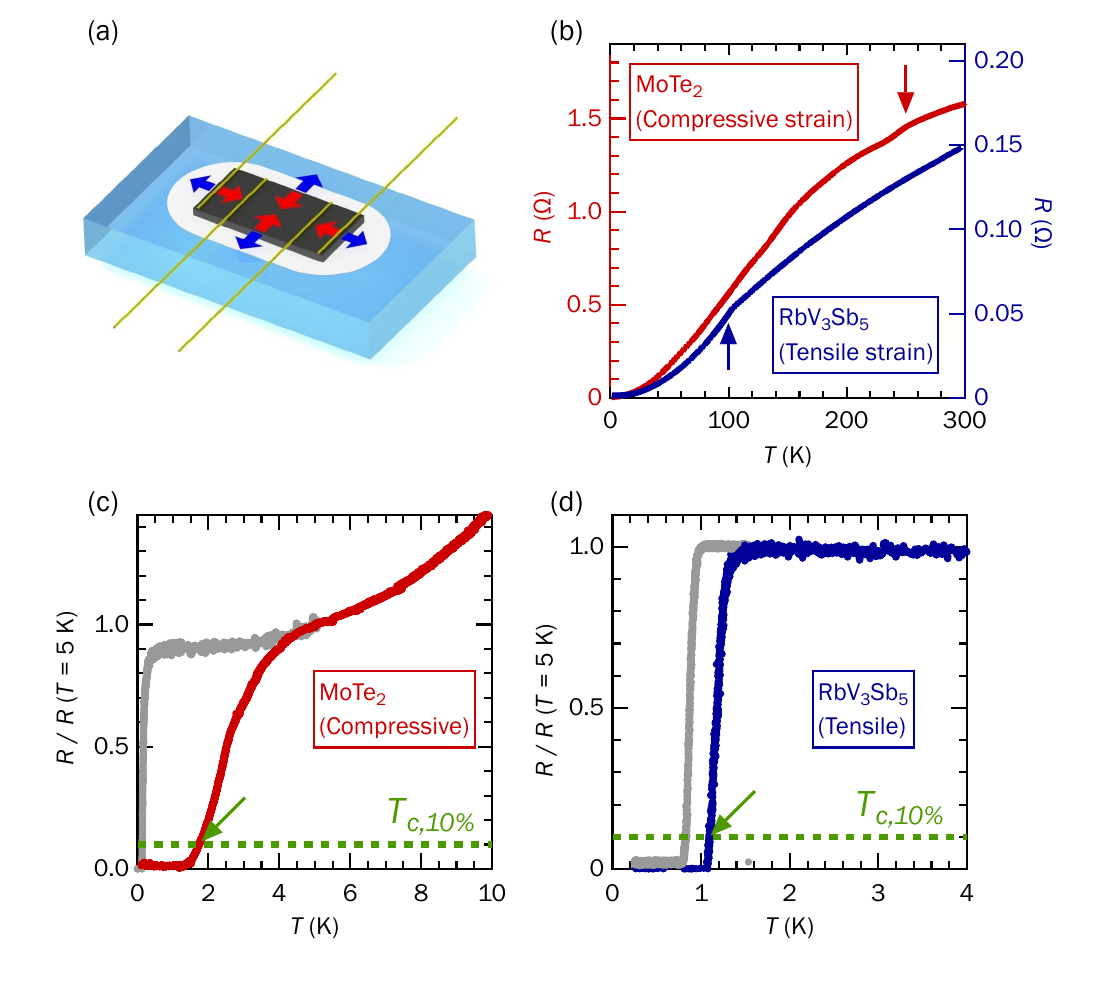}}\
              \caption{\label{fig1} 
              (a) Schematics of the device for biaxially straining a sample (black) by utilizing the differential thermal expansion between the sample and the substrate (light blue). The sample is mechanically coupled to the substrate with a suitable adhesive (white). The device is equipped with four gold wires for magnetotransport studies. Both the compressive and tensile strains can be realized by a careful selection of the substrate. (b) Temperature dependence of the electrical resistance ($R(T)$) of strained MoTe$_2$ (red) and RbV$_3$Sb$_5$ (blue) at zero magnetic field. (c) Normalized $R(T)$ curves showing the superconducting transitions of free-standing MoTe$_2$ (grey) and MoTe$_2$ under a compressive strain (red). (d) Normalized $R(T)$ curves showing the superconducting transitions of free-standing RbV$_3$Sb$_5$ and RbV$_3$Sb$_5$ under a tensile strain (blue).  The horizontal dashed green lines indicate the `$10~\%$ criterion' for the definition of $T_c$.
              }
\end{figure}

Single crystals of MoTe$_2$ and RbV$_3$Sb$_5$ were prepared by the self-flux method as described elsewhere~\cite{Hu2019,Wang2023}. These single crystals were used in both free-standing and biaxial strain measurements. The dimensions ($L~\times~W~\times~H$) of MoTe$_2$ and RbV$_3$Sb$_5$ for the biaxial strain devices are 818~$\times$~211~$\times$~14~$\mu$m$^3$ and 804~$\times$~255~$\times$~11~$\mu$m$^3$, respectively. The MoTe$_2$ sample was attached to a polycarbonate substrate with Cyanoacrylate (CN) adhesives (Tokyo Measuring Instruments Lab Co.,~Ltd.). The RbV$_3$Sb$_5$ sample was glued on a sapphire substrate 
with Threebond 2086M two-component epoxy resin adhesive (ThreeBond Holdings Co., Ltd.). A schematic setup is shown in Fig.~\ref{fig1}(a). Strain gauges (Tokyo Measuring Instruments Lab Co.,~Ltd.) were used to characterize the compressive (tensile) strain induced on MoTe$_2$ (RbV$_3$Sb$_5$) at cryogenic temperatures. The biaxial strain induced on MoTe$_2$ and RbV$_3$Sb$_5$ were found to be $-1.06$~\% and $+1.16$~\%, respectively, with the positive sign indicating tensile strain as a convention. See Supplementary Material for more details. The temperature-dependent resistance was measured by the standard four-probe method. All low-field data were collected with Lakeshore AC resistance bridge in a Bluefors dilution refrigerator down to 10~mK. All high-field data were collected with a customized lock-in detection unit in a Helium-3 cryostat with a pulsed field magnet at The National High Magnetic Field Laboratory's Pulsed Field Facility at Los Alamos National Laboratory.  

To demonstrate the versatility of the biaxial strain methodology, we have applied compressive and tensile strain on MoTe$_2$ and RbV$_3$Sb$_5$, respectively. Both MoTe$_2$ and RbV$_3$Sb$_5$ are novel superconductors currently being investigated intensively~\cite{Chen2016,Rhodes2017,Qi2016,Lam2023,Yip2023,guguchia2023,Ortiz2019,Yin2021,Cho2021,Zhu2022,Du2022,Shrestha2023,Wang2023PRB,Wang2023,Frassineti2023,wang2023b}. Figure~\ref{fig1}(a) shows the schematic of the biaxial strain device. To prepare this device, a thin flake cleaved from a bulk single crystal is mechanically coupled to a piece of insulating substrate, which is $\sim$10-mm-thick, with a suitable adhesive at room temperature. As the device is cooled towards 0~K, the differential thermal expansion causes the substrate to exert biaxial strain on the thin flake sample. For the case of MoTe$_2$ on polycarbonate, the resultant strain at the lowest attainable temperature is compressive, while for RbV$_3$Sb$_5$ on sapphire, tensile strain is induced.

The gold wires attached to the thin flake enable electrical transport measurements of the strained samples. Figure~\ref{fig1}(b) shows the temperature dependence of resistance, $R(T)$, of MoTe$_2$ (red) and RbV$_3$Sb$_5$ (blue) measured with the simple device architecture in Fig.~\ref{fig1}(a), indicating high signal quality. The anomaly at $T_s\approx 250$~K for MoTe$_2$ (red arrow) can be attributed to the $1T'$ to $T_{d}$ structural phase transition. On the other hand, RbV$_3$Sb$_5$ exhibits a charge density wave (CDW) transition at $T_{\rm CDW}\approx 100$~K. These phase transition temperatures are consistent with the corresponding free-standing samples in previous studies~\cite{Qi2016,Hu2019,Hu2020,Yin2021,Neupert2021,Wang2023}. This is expected since the induced biaxial strain only fully develops at the 0~K limit and is not effective at elevated temperatures. 
\begin{figure}[!t]\centering
       \resizebox{8.5cm}{!}{
              \includegraphics{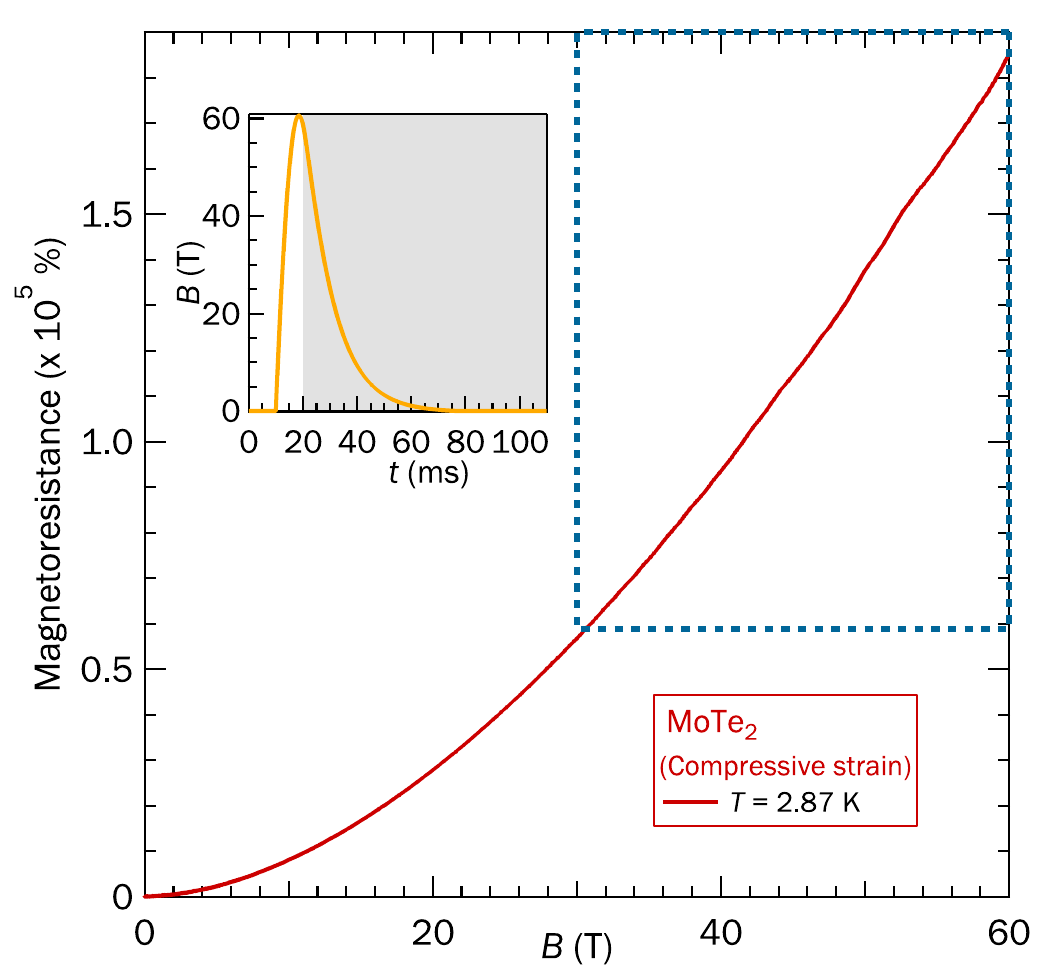}}  
              
              \caption{\label{fig2} 
              Magnetoresistance of strained MoTe$_2$ at 2.87~K. The high field data in the dashed rectangle was used for SdH oscillation analysis. The top inset illustrates a representative magnetic-field pulse profile up to 60~T. The data from the descending field (shaded area) are used in the analysis. The field direction is perpendicular to the $ab$-plane of the sample.}
\end{figure}

To explore the effect of the biaxial strain at low temperatures, we turn to another quantum state, namely superconductivity. Figures~\ref{fig1}(c) and~\ref{fig1}(d) compare the normalized $R(T)$ for the strained MoTe$_2$ (red) and strained RbV$_3$Sb$_5$ (blue) against their free-standing counterparts (grey curves). Bench-marking the superconducting critical temperature ($T_c$) with the `10~\% criterion', where the resistance drops to 10~\% of the normal state resistance, the application of strain boosts the $T_c$ from 0.13~K to 1.7~K for MoTe$_2$, and from 0.82~K to 1.1~K for RbV$_3$Sb$_5$. The large enhancement of superconductivity in MoTe$_2$ due to biaxial compressive strain has also been pointed out recently in Ref.~[\onlinecite{Yip2023}]. Thus, the biaxial device is particularly powerful for exploring the ground state of layered quantum materials, or for probing them near 0~K, prompting the investigation of the electronic structure of MoTe$_2$ and RbV$_3$Sb$_5$ in higher fields using a pulsed magnet.

The main panel of Fig.~2 shows the magnetoresistance of the strained MoTe$_2$ device measured up to 60~T at $2.87$~K. The field dependence of the resistance has been corrected with a field-independent offset, as detailed in Supplementary Material. The magnetic field was applied perpendicular to the $ab$-plane of the sample. A typical magnetic-field pulse profile is shown in the inset and all data analyzed in this manuscript are from the portion collected with the descending field, indicated by the shaded region. The magnetoresistance curve in the main panel shows a nearly $B^2$ behaviour, and the resistance is non-saturating even at 60~T. A non-saturating, nearly quadratic-in-field magnetoresistance has also been observed in the free-standing MoTe$_2$ as well as our previous strained sample~\cite{Zhou2016,Hu2020,Yip2023}. At 60~T, the magnitude of the magnetoresistance (MR) reaches $\sim$185,000~\%. For $B\geq30$~T, the magnetoresistance exhibits ripples (data enclosed in the dashed rectangle), which are in fact SdH oscillations. 
\begin{figure}[!t]\centering
       \resizebox{8.5cm}{!}{
              \includegraphics{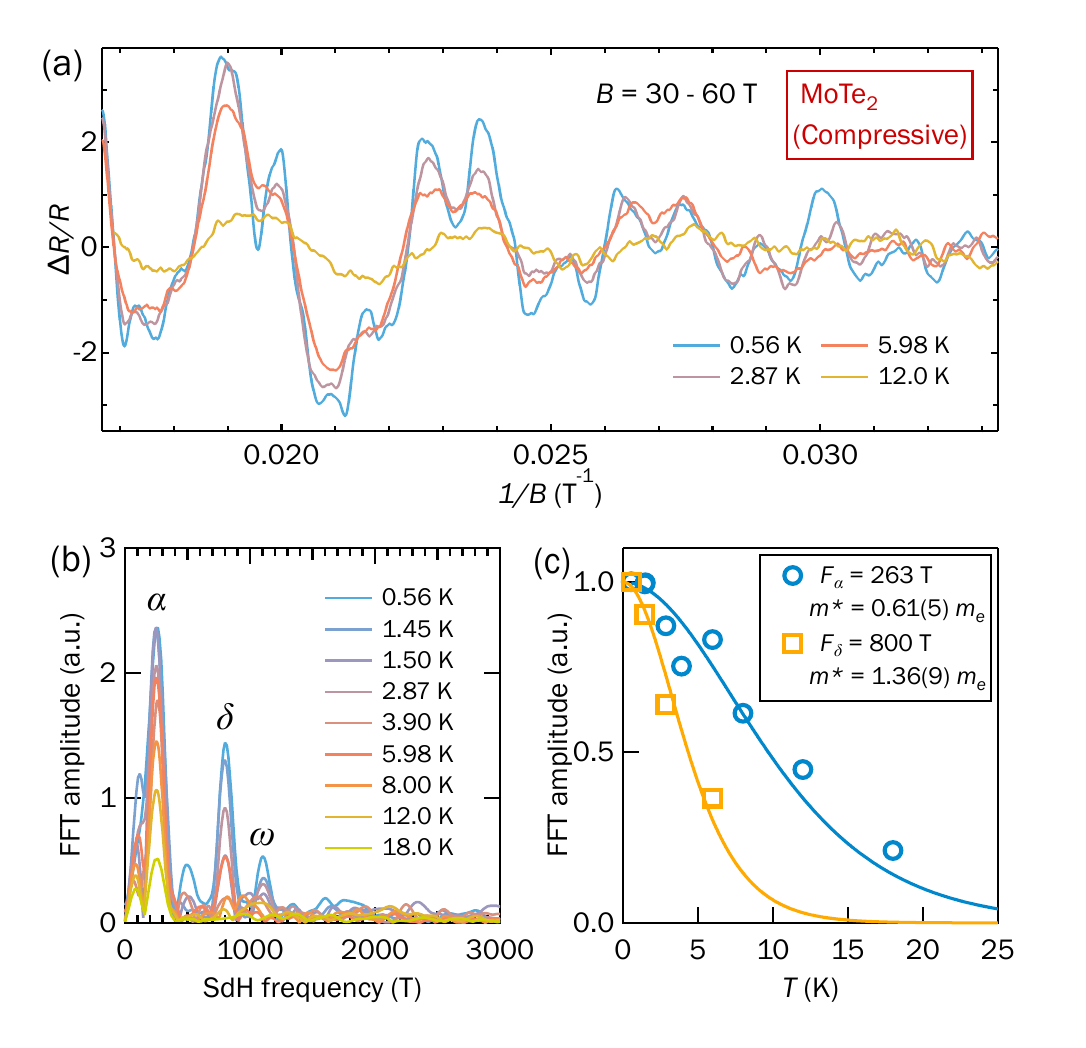}}                				
              \caption{\label{fig3} (a) Representative SdH oscillations of strained MoTe$_2$ at different temperatures after removing the magnetoresistance background. (b) FFT spectra at various temperatures ranging from 0.56~K to 18~K, revealing clear peaks at $\alpha\sim263$~T, $\delta\sim800$~T, and $\omega\sim1103$~T. (c) Temperature dependence of the SdH amplitudes for $F_\alpha$ (open circles) and $F_\delta$ (open squares). The solid curves are the fittings by Lifshitz-Kosevich (LK) formula (Eqn.~\ref{eq_LK}) and extracted effective masses are shown.}
\end{figure}

\begin{figure}[!t]\centering
       \resizebox{8.5cm}{!}{
              \includegraphics{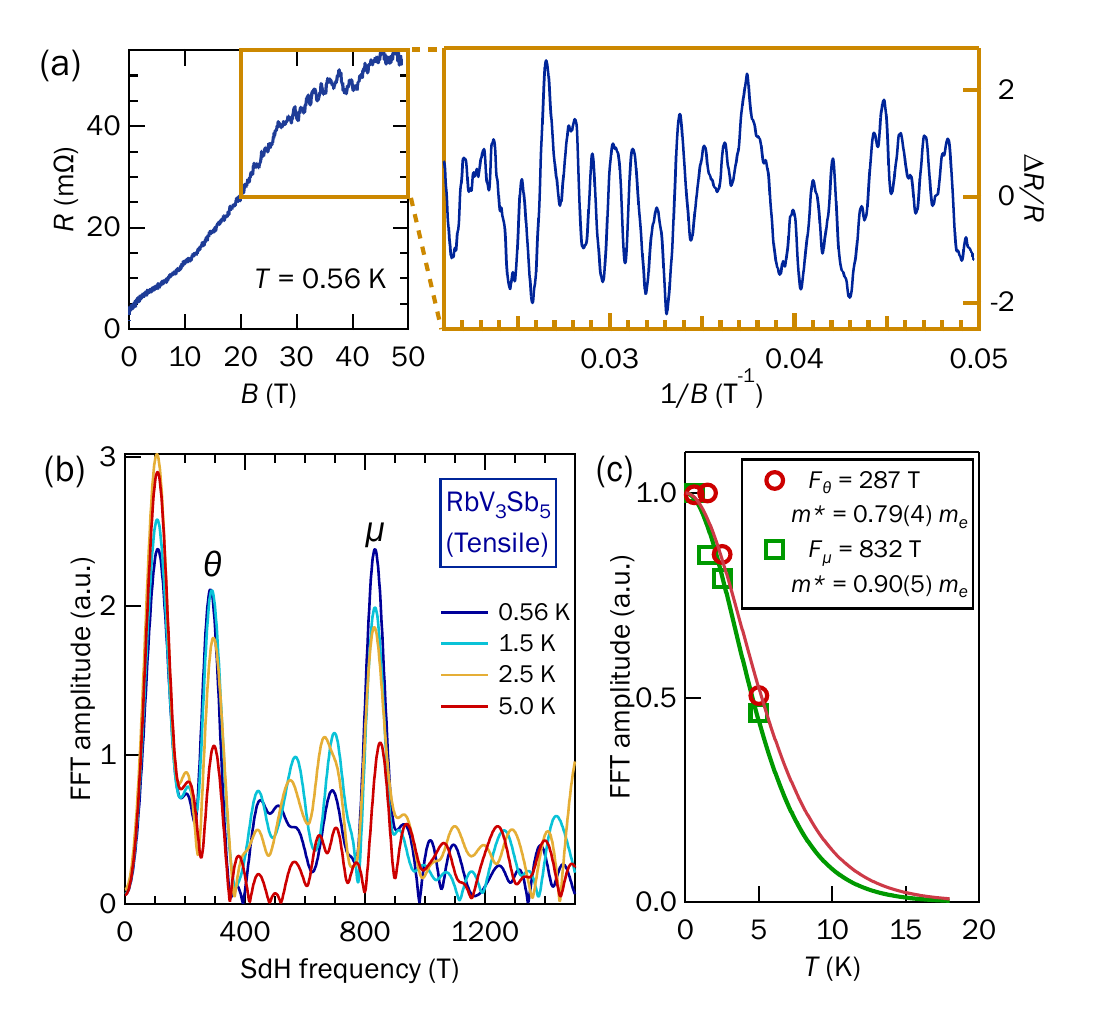}}                				
              \caption{\label{fig4} (a) Magnetic field dependence of the electrical resistance for strained RbV$_{3}$Sb$_{5}$ at $T=0.56$~K. A close-up view of SdH oscillation signals are shown on the right panel after background removal. (b) FFT spectra at various temperatures. (c)  Temperature dependence of the SdH amplitudes for $F_{\theta}\sim$287~T (open circles) and $F_{\mu}\sim$832~T (open squares). The solid curves are the corresponding Lifshitz–Kosevich fittings (Eqn.~\ref{eq_LK}), and the resultant effective masses are shown in the figure.}
\end{figure}
After removing the large magnetoresistance background, clear SdH oscillations in strained MoTe$_{2}$ are immediately visible, with representative curves displayed in Fig.~\ref{fig3}(a). The SdH oscillations deteriorate with increasing temperature, as expected. The fast Fourier Transform (FFT) was performed between $B=30$~T and $B=60$~T for all temperatures to quantify these oscillations. The FFT spectra at different temperatures are plotted in Fig.~\ref{fig3}(b). Three SdH frequencies, $F_{\alpha}$, $F_{\delta}$, and $F_{\omega}$, are well resolved at 263~T, 800~T, and 1103~T, respectively. $F_{\alpha}$ is consistent with an electron pocket observed in previous studies on free-standing samples~\cite{Qi2016,Rhodes2017,Hu2020,Liu2020} and also on a biaxially strained MoTe$_2$~\cite{Yip2023}. $F_{\delta}$, on the contrary, is likely related to the SdH frequency of 758~T reported for the free-standing MoTe$_2$~\cite{Hu2020}. In addtion, $F_{\omega}$ is close to the sum of $F_{\alpha}$ and $F_{\delta}$. To extract more information about these Fermi pockets, we analyze the SdH oscillations based on the thermal damping factor $R_T$ of the Lifshitz-Kosevich (LK) theory, which can be expressed as:
\begin{equation}
    R_{T}=\frac{14.693m^{*}T/B}{\sinh{(14.693m^{*}T/B})}
    \label{eq_LK}.
\end{equation}
From Eqn.~(\ref{eq_LK}), we can extract the effective mass of the quasiparticles ($m^{*}$) on extremal cyclotron orbits of the Fermi sheets perpendicular to the magnetic field direction. The effective mass of $\alpha$ and $\delta$ is found to be 0.61(5) and 1.36(9) times the bare electron mass $m_{e}$, respectively. The effective mass of $\omega$ cannot be extracted because the FFT amplitudes are too weak. The $m^{*}$ of $\alpha$ is similar to that in the free-standing MoTe$_{2}$~\cite{Rhodes2017,Liu2020,Hu2020}. However, the $m^{*}$ of $\delta$ is significantly different from what is observed in free-standing MoTe$_{2}$ where $m^{*}=1.99(6)m_{e}$~\cite{Rhodes2017,Liu2020,Hu2020}. The above observations reveal that the $\delta$ pocket is much more sensitive to biaxial strain tuning than the $\alpha$ pocket. Considering the drastic enhancement in $T_c$ by biaxial strain tuning, it is quite likely that the $\delta$ pocket plays a major role in this effect.

Next, we switch to the tensile-strained RbV$_{3}$Sb$_{5}$ and investigate its electronic properties. Figure~\ref{fig4}(a) displays the $R(B)$ data of strained RbV$_{3}$Sb$_{5}$ at 0.56~K. The $R(B)$ curve shows a monotonically increasing magnetoresistance which reaches $\sim$2800~\% at 50~T, indicating reasonably high sample quality~\cite{Wang2023,Wang2023PRB}. The $R(B)$ background is removed by using a polynomial from 20~T to 50~T and the SdH oscillations are extracted. The resultant oscillatory signals are clearly visible and shown in the right panel of Fig.~\ref{fig4}(a). The SdH oscillations at various temperatures were analyzed by FFT and the spectra are plotted in Fig.~\ref{fig4}(b). Here, we focus on SdH frequencies above 200~T where FFT analysis at this field window (20-50~T) works well. To resolve frequencies lower than 200~T, good SdH signals down to lower fields are necessary.
Two SdH frequencies, labelled as $F_{\theta}$ and $F_{\mu}$, were resolved at 287~T and 832~T in the spectra, respectively. Since the FFT spectra reported earlier for free-standing samples are complicated~\cite{Yin2021,Shrestha2023,Wang2023}, we proceed by performing the LK analysis of $F_{\theta}$ and $F_{\mu}$ to look for correspondence between them and the SdH frequencies in free-standing RbV$_3$Sb$_5$. 

The effective mass analysis of $F_{\theta}$ and $F_{\mu}$ suggests an intricate dependence on biaxial strain. Figure~\ref{fig4}(c) shows the LK fittings on the SdH amplitudes of $F_{\theta}$ and $F_{\mu}$ using Eqn.~\ref{eq_LK}. Surprisingly, we find that the $m^{*}$ values for $F_{\theta}$ and $F_{\mu}$ are $0.79(4)m_{e}$ and $0.90(5)m_{e}$, respectively, which are substantially larger than all $m^*$ values reported in free-standing RbV$_3$Sb$_5$~\cite{Yin2021,Wang2023,Wang2023PRB,Shrestha2023}. The enhanced $m^*$ is likely attributable to increased charge fluctuations, resulting from the weakening of the CDW order when the $ab$-plane is enlarged by the biaxial strain. Comparisons between uniaxial strain and hydrostatic pressure effects on the sister compound CsV$_3$Sb$_5$ have established that the shortening of the out-of-plane lattice constant $c$ is the primary parameter driving the initial increase of $T_c$ and the decrease of $T_{\rm CDW}$~\cite{qian2021,Yang2023}. The decrease of $c$ in uniaxial experiments is due to the Poisson effect. Thus, a decrease of $c$ can be similarly expected when biaxial tensile strain acts on RbV$_3$Sb$_5$. Indeed, the observed enhancement of $T_c$ under biaxial strain lends support to this hypothesis, and the increased charge fluctuations can be felt by the quasiparticles as the CDW phase in RbV$_3$Sb$_5$ is progressively suppressed, resulting in enhanced $m^*$ values. However, other possible scenarios which can lead to enhancement in $T_{c}$, such as the change in charge carrier density induced by biaxial strain, should also be noted. Further Hall effect and detailed SdH measurements would shed light on the role of biaxial strain on $T_{c}$ and $m^{*}$ in RbV$_3$Sb$_5$.


Our pilot magnetotransport measurements on MoTe$_2$ and RbV$_3$Sb$_5$ up to 60~T clearly demonstrate the feasibility of probing the electronic structures of layered materials tuned by a biaxial strain. While we are mindful of the drawbacks that the strain only fully develops at the zero temperature limit, the methodology offers an elegant and simple way to perturb and study the ground state of quantum materials. The small footprint of the biaxial device further facilitates the integration with other pulsed-field magnets, and the flexible wiring arrangement can enable other high-field experiments. For instance, we do not foresee any technical barriers in measuring the Hall effect -- regardless of whether it is classical or quantum, normal or anomalous -- up to 100~T while simultaneously tuning the system by biaxial strain.

Hydrostatic pressure has been extremely successful in materials tuning. However, it is challenging to integrate pressure devices into a pulsed-field magnet. This is because pressure devices are usually bulky and complicated, requiring elaborate effort to prepare, and they contain many metallic parts. The present methodology of utilizing the thermal expansion mismatch of an insulating substrate circumvents all these issues. Finally, while hydrostatic pressure is always compressive, biaxial strain can be either compressive or tensile with a careful selection of the substrate, as has been demonstrated in this manuscript. Therefore, the usage of biaxial strain as a tuning parameter offers the possibility of addressing unique challenges in materials research, and its integration with pulsed magnetic field holds great promise for advancing our knowledge in the field of quantum materials.

To summarize, we have applied biaxial strain using a convenient method on superconductors MoTe$_{2}$ and RbV$_{3}$Sb$_{5}$, and conducted magnetotransport measurements in a pulsed magnetic field. $T_c$ of MoTe$_{2}$ increases substantially from 0.13~K to 1.7~K upon the application of compressive biaxial strain by polycarbonate substrate. On the other hand, a sapphire plate is used to apply tensile strain on RbV$_{3}$Sb$_{5}$ and its $T_c$ has been raised from 0.82~K to 1.1~K.  We have observed quantum oscillations in magnetoresistance, the SdH effect, in both devices. In compressive-strained MoTe$_2$, one of the SdH frequencies, $F_{\delta}$, has shown a decrease in effective mass $m^*$ compared to the free-standing sample, which could be relevant to the observed $T_c$ enhancement under strain. In tensile-strained RbV$_{3}$Sb$_{5}$, we observe substantial increases in the $m^*$ values associated with $F_{\theta}$ and $F_{\mu}$, implying that biaxial strain is effective in tuning electron correlations in this Kagome system. This study illustrates that the combination of the biaxial strain and pulsed magnetic field provides a powerful platform for tuning and investigating topical materials under unprecedented conditions.

\section*{Supplementary Material}

The details for the determination of biaxial strain induced in MoTe$_2$ and RbV$_3$Sb$_5$, and for the corrections of magnetoresistance in Fig.~\ref{fig2} are described in the Supplementary Material.

\begin{acknowledgments}

We thank Wing Shing Chow and Jiayu Zeng for the preliminary development of the methodology for biaxial strain application, and Chun Wai Tsang and Lut Yue Tai for advice on crystal growth. We acknowledge financial support from Research Grants Council of Hong Kong (GRF/14300722, GRF/14301020 and A-CUHK402/19), CUHK Direct Grant (4053463, 4053528, 4053408 and 4053461), the National Natural Science Foundation of China (12104384). A portion of this work was performed at the National High Magnetic Field Laboratory, which is supported by National Science Foundation Cooperative Agreement No. DMR-2128556, the State of Florida and the U.S. Department of Energy.\\

\noindent$^\dagger$K.Y.Y. and L.W. contributed equally to this work.

\end{acknowledgments}

\section*{AUTHOR DECLARATIONS
}
\subsection*{{Conflict of Interest
}}
The authors have no conflicts to disclose.

\section*{Data Availability Statement}
The data that support the findings of
this study are available from the
corresponding author upon reasonable
request.

\providecommand{\noopsort}[1]{}\providecommand{\singleletter}[1]{#1}%

\end{document}